\documentstyle[aps,multicol,epsfig]{revtex}
\title{Quantum copying: Fundamental inequalities}
\author{
M. Hillery$^{1}$ and V. Bu\v{z}ek$^{2,3}$}
\address{
$^{1}$Department
of Physics and Astronomy, Hunter College, CUNY,
695 Park Avenue, New York, NY 10021, USA\newline
$^{2}$ Optics Section, The Blackett Laboratory,
Imperial College, London SW7 2BZ,  England\newline
$^{3}$ Institute of Physics, Slovak Academy of Sciences, Dubravsk\'a
cesta 9, 842 28 Bratislava, Slovakia
}
\date{January 20, 1997}

\begin{document}

\maketitle
\begin{abstract}
How well one can copy an arbitrary qubit? To answer this question we
consider two arbitrary vectors in a two-dimensional state space and an
abstract copying transformation which will copy these two vectors. If the
vectors
are orthogonal, then perfect copies can be made. If they are not, then
errors will be introduced. The size of the error depends on the inner
product of the two original vectors. We derive a lower bound for the amount
of noise induced by  quantum copying. We examine both copying transformations
which produce one copy and transformations which produce many, and show that
the quality of each copy decreases as the number of copies increases.
\end{abstract}
\begin{multicols}{2}
\narrowtext

\section{INTRODUCTION}

One of the greatest differences between classical and quantum information is
that while classical information can be copied perfectly, quantum cannot. In
particular, we cannot create a duplicate of an {\em arbitrary} quantum bit (%
{\em qubit}) \cite{Ekert1} without destroying the original. This follows
from the {\em no--cloning theorem} of Wootters and Zurek \cite{Wootters1}
(see also \cite{Diekes,Barnum}). There are many consequences of this
theorem. For example, if one has a string of qubits which one would
like to process in more than one way, it represents a serious limitation.
With a string of classical bits, one could simply copy the string and
process the original one way and the copy another.  Quantum mechanically
this is impossible.  On the other hand, the fact that information cannot
be copied is sometimes an advantage.
One can view the impossibility of quantum copying as one of the main reasons
why quantum cryptography works. In a quantum cryptographic system \cite
{Ekert2,Bennett1} qubits are exchanged between a sender (Alice) and a
receiver (Bob) in such a way that the presence of an eavesdropper (Eve)
can be detected. If quantum copying were possible the eavesdropper could
simply copy the qubits which Alice is sending to Bob, and they would
not be able to detect this procedure. This would leave the eavesdropper with
a perfect record of their communication. The fact that quantum information
cannot be copied rules out this possibility.

Even though one cannot copy quantum information perfectly, it is useful to
know how well one can do. One would like to know to what extent it is
possible to split the information in a given qubit among several others.
In addition, if it is possible to make close to
perfect copies quantum cryptographic schemes could still be at risk
\cite{Gisin}. Finally, quantum copying
can become essential in storage and retrieval of information in
quantum computers \cite{DiVincenzo}.

In our previous paper we examined several possible quantum copying
machines\footnote{In what follows we will use a shorthand ``copying machine''
for a particular unitary transformation applied to the original particle.
We do this having in mind that copying unitary transformations	under
consideration  can be realized in terms of a sequence of logical gates.}
and studied how they would perform copying a single {\em arbitrary} qubit
\cite{Buzek1}. The copier proposed in Wootters and Zurek's paper \cite
{Wootters1} on quantum cloning copies two orthogonal states perfectly but
introduces errors when superpositions of these states are copied. A second
copying machine, which we called the universal quantum copying machine,
copies all input states to the same accuracy, and, on average, its
performance is much better than that of the Wootters-Zurek machine. Here we
would like to establish some fundamental limits on how well quantum states
can be copied by considering the following problem. Suppose we have two
arbitrary vectors in a two-dimensional state space and we want to build a
machine which will copy these two vectors. How well we can do? If the
vectors are orthogonal, then perfect copies can be made. If they are not,
then, as we shall show, errors will be introduced. The amount of error
depends on the inner product of the two original vectors. This problem is
relevant to the global problem of copying an arbitrary qubit. If one
has a lower bound for the amount of noise which must be introduced for the
two-state problem, then the best one can do in the general case is the
maximum of this lower bound over all pairs of states. Thus we can get a
lower bound for the amount of noise induced by a quantum copying machine.

The approach which we use here has the advantage that it allows us to
consider more general problems than simply producing a single copy of
an arbitrary qubit.  We are able to find a lower bound for the noise
which is introduced when $n$ copies of a qubit are produced simultaneously,
and determine how the noise depends on $n$.  In addition, even though
our discussion is phrased in terms of qubits, which are two-level
systems, our results are more general; the limitations we find on quantum
copying apply to systems of arbitrary dimension, because our arguments
are completely independent of the dimension of the Hilbert space in
which the vectors to be copied lie.  Therefore, if one is trying to
copy an $n$-level system, for example several qubits in an entangled
state, then the amount of noise introduced by the copying process
must be greater than the lower bounds which are given here.

\section{TWO-STATE PROBLEM}

Suppose we have two states $| s_1\rangle_a$ and $| s_2\rangle_a$, in a
two-dimensional state space which we would like to copy. If the initial
state of the copy machine is $| Q\rangle_x$, then the action of the copying
machine on our two vectors can be expressed as
\begin{eqnarray}
| s_j\rangle_a | Q\rangle_x \rightarrow |\Psi_{j}\rangle_{abx}= |
s_j\rangle_a | s_j\rangle_b | Q_j\rangle_x + | \Phi_j\rangle_{abx},
\label{e1}
\end{eqnarray}
where $j=1,2$. In our analysis we do not specify the {\em in}-state of the
copy mode (this possible eavesdropper's mode we denote as the $b$-mode). We
only require that it is the same for all inputs into the $a$ mode, and that
it is normalized to unity. In Eq.(\ref{e1}) we have expressed the full
output state of the copy machine as the sum of two parts, the first
representing the ideal output state and the second what is left over. The
two parts can be expressed in terms of the projection onto the two mode
state $|s_{j}\rangle_{a} |s_{j}\rangle_{b}$  as
\begin{equation}
|\Gamma_j\rangle_{abx}\equiv
|s_{j}\rangle_{a}|s_{j}\rangle_{b}|Q_{j}\rangle_{x} =
P_{j}|\Psi_{j}\rangle_{abx};  \label{e1a}
\end{equation}
\begin{equation}
|\Phi\rangle_{abx}\equiv (I-P_{j})|\Psi\rangle_{abx},  \label{e1b}
\end{equation}
where the projectors $P_j$ are defined as
\begin{equation}
P_{j}=(|s_{j}\rangle\langle s_{j} |)_{a}\otimes (|s_{j}\rangle\langle s_{j}
|)_{b}.  \label{e1c}
\end{equation}
This definition implies that
\begin{eqnarray}
_{abx}\langle \Gamma_j|\Phi_j\rangle_{abx}=0; ~~~ j=1,2.  \label{e2}
\end{eqnarray}
In addition we also assume that the initial quantum-copying machine state is
normalized to unity, i.e. $_x\langle Q| Q\rangle_x =1$. In order to produce
good copies we want to make the norms $\|Q_1\|$ and $\|Q_2\|$ as large as
possible and $\|\Phi_1\|$ and $\|\Phi_2\|$, which represent the size of the
errors, as small as possible. The norm of the state vector $| A\rangle$ is
defined as $\| A\| =(\langle A| A\rangle)^{1/2}$.

The copying machine can be represented as a unitary operator and this
unitarity impose constraints on the transformations shown in Eq.(\ref{e1}).
In particular, we have that
\begin{eqnarray}
1=\| Q_j\|^2 + \|\Phi_j\|^2, ~~~ j=1,2	\label{e3}
\end{eqnarray}
and
\begin{eqnarray}
z & = & z^2\, _x\langle Q_1 | Q_2\rangle_x +\, _{abx}\langle \Gamma_1
|\Phi_2\rangle_{abx}  \nonumber \\
& + &\, _{abx}\langle \Phi_1 |\Gamma_2\rangle_{abx} +\, _{abx}\langle \Phi_1
|\Phi_2\rangle_{abx},  \label{e4}
\end{eqnarray}
where $z=\, _a\langle s_1 | s_2\rangle_a$. We note that in derivation of Eq.(%
\ref{e4}) we have utilized the fact that the {\em in}-state of the copy mode
is normalized to unity. From these equations it is possible to derive a
number of inequalities which restrict the behaviour of the copy machine. We
shall begin with the strongest restriction, which is relatively difficult to
work with, and then we proceed to weaker ones which are more transparent.

Let us first find an upper bound on $|_{abx}\langle \Gamma_1
|\Phi_2\rangle_{abx}|$ and $|_{abx}\langle \Phi_1 |\Gamma_2\rangle_{abx}|$.
We begin by expressing $|\Gamma_1\rangle_{abx}$ as
\begin{equation}
|\Gamma_1\rangle_{abx} = P_{2} |\Gamma_1\rangle_{abx} + |\Gamma_1
^{\prime}\rangle_{abx}.  \label{e5}
\end{equation}
where $|\Gamma_{1}^{\prime}\rangle_{abx} = (I-P_{2}) |\Gamma_{1}\rangle_{abx}
$. The two states on the right hand side of Eq.(\ref{e5}) are orthogonal
which implies that
\begin{equation}
\eta_{11}=\eta_{11}|z|^{4}+\|\Gamma_{1}^{\prime}\|^{2},  \label{e6}
\end{equation}
where $\eta_{ij}=\, _x\langle Q_i | Q_j\rangle_x $, so that
\begin{equation}
\|\Gamma_1 ^{\prime}\| =\left[\eta_{11}(1 - |z|^4)\right]^{1/2},  \label{e7}
\end{equation}
Similarly, if we express $|\Gamma_{2}\rangle_{abx}$ as
\begin{equation}
|\Gamma_2\rangle_{abx} = P_{1} |\Gamma_2\rangle_{abx} + |\Gamma_2
^{\prime}\rangle_{abx}.  \label{e8}
\end{equation}
where $|\Gamma_{2}^{\prime}\rangle_{abx}=(I-P_{1}) |\Gamma_{2}\rangle_{abx}$%
, we find
\begin{equation}
\|\Gamma_2 ^{\prime}\| =\left[\eta_{22}(1 - |z|^4)\right]^{1/2}.  \label{e9}
\end{equation}
Because $P_{2} | \Phi_2\rangle_{abx} = 0$ we have that
\begin{eqnarray}
\left|\, _{abx}\langle \Phi_2 | \Gamma_1\rangle_{abx} \right|= \left|\,
_{abx}\langle \Phi_2 | \Gamma_1^{\prime}\rangle_{abx} \right| \leq
\|\Gamma_1^{\prime}\| \cdot \|\Phi_2\|	\label{e10} \\
= \left( 1- \eta_{22}\right)^{1/2} \left[\eta_{11}(1 - |z|^4)\right]^{1/2}
\nonumber
\end{eqnarray}
and similarly
\begin{eqnarray}
\left|\, _{abx}\langle \Phi_1 | \Gamma_2\rangle_{abx} \right| \leq \left( 1-
\eta_{11}\right)^{1/2} \left[\eta_{22}(1 - |z|^4)\right]^{1/2}.  \label{e11}
\end{eqnarray}

We can now take these results and insert them into Eq.(\ref{e4}). This gives
us
\begin{eqnarray}
|z|\leq |z|^2\, |\eta_{12}|+ \left(1-\eta_{11}\right)^{1/2}
\left(1-\eta_{22}\right)^{1/2}	\nonumber \\
+(1-|z|^{4})^{1/2} \left[\eta_{11}^{1/2}(1-\eta_{22})^{1/2}
+\eta_{22}^{1/2}(1-\eta_{11})^{1/2}\right].  \label{e12}
\end{eqnarray}
For a given value of $|z|$ this inequality restricts the values of $\| Q_1\|$%
, $\| Q_2\|$, and $|\eta_{12}|=|\langle Q_2|Q_1\rangle|$. It defines a
region in a 3-dimensional parametric space in which the values of the
parameters can lie. For $|z|\neq 0$ this region does not include the line $%
\| Q_1\| = \| Q_2\| = 1$ which implies that perfect copying is impossible.
It is only for $| z| =0$, i.e., $|s_1\rangle$ and $|s_2\rangle$ are mutually
orthogonal, that we can have $\| Q_1\| =\| Q_2\| =1$ which implies
error-free copying.

In order to simplify these results we use the Schwarz inequality from which
it follows that
\begin{eqnarray}
|\eta_{12}| \leq \| Q_1\| \, \| Q_2\| = (\eta_{11}\eta_{22})^{1/2}.
\label{e13}
\end{eqnarray}
This last inequality allows us to rewrite the right-hand side of the
relation (\ref{e12}) in terms of only two parameters, $\eta_{11}$ and $%
\eta_{22}$. It is useful to express the resulting inequality  in terms of
the size of the errors. We introduce the quantities $X_{j}=(1-%
\eta_{jj})^{1/2}= \|\Phi_{j}\|$ (for $j=1,2$) which are associated with the
amount of noise induced by copying the vectors $|s_{j}\rangle_{a}$. In
particular, the smaller $X_1$ and $X_2$ are the better is the copying
procedure, and in the limit $X_j\rightarrow 0$ two perfect copies $|
s_j\rangle_a$ and $| s_j\rangle_b$ of the initial state $| s_j\rangle_a$ are
obtained at the output of the copying machine. If we now express the
inequality which follows from Eqs.(\ref{e12}) and (\ref{e13}) in terms of $%
X_{1}$ and $X_{2}$ we have
\begin{eqnarray}
|z|&\leq &|z|^{2}(1-X_{1}^{2})^{1/2}(1-X_{2}^{2})^{1/2} +X_{1}X_{2}
\nonumber \\
& & +(1-|z|^{4})^{1/2}[(1-X_{1}^{2})^{1/2}X_{2} +(1-X_{2}^{2})^{1/2}X_{1}].
\label{e16a}
\end{eqnarray}
It is easiest to understand the implications of Eq.(\ref{e16a}) if we look
at particular cases.

{\bf (A)} Let us first suppose that $X_1= \| \Phi_1\|=0$, i.e. $| s_1\rangle$
is copied perfectly, which implies that $\|Q_1\|=1$. From Eq.(\ref{e16a}) we
find
\begin{eqnarray}
|z|\leq |z|^2\, \left( 1-X_2^2\right)^{1/2} + \left(1-|z|^4\right)^{1/2} X_2,
\label{e18}
\end{eqnarray}
which in turn implies that
\begin{eqnarray}
X_2\geq |z| \left(1-|z|^2\right)^{1/2} \left[\left(1+|z|^2\right)^{1/2} -
|z|\right].  \label{e19}
\end{eqnarray}
Therefore, if $| s_1\rangle$ is copied perfectly, then $\| \Phi_2\|$, which
represents the size of the error made in copying $| s_2\rangle$, must be
{\em at least} as large as the right-hand side of Eq.(\ref{e19}). For small $%
|z|$ the right-hand side of this inequality is approximately $|z|$. We note
that the maximum value of the lower bound on the error $X_2$ given by the
right-hand side of Eq.(\ref{e19}) is equal to $(2/27)^{1/2}\simeq 0.272$ and
is obtained for $|z|=1/\sqrt{3}\simeq 0.577$.

{\bf (B)} Let us now consider the case $X_1=X_2=X$, i.e. equal errors in
both copies. Making use of Eq. (\ref{e16a}) we then have that
\begin{eqnarray}
|z|\leq |z|^2\, \left( 1-X^2\right) + X^2 +2 X
\left[\left(1-|z|^4\right)
\left(1-X^2\right)\right]^{1/2}  \label{e20}
\end{eqnarray}
which implies that
\begin{eqnarray}
X\geq \left[
\frac{r_1- 2 r_2^{1/2}}{r_3}\right]^{1/2}
\label{e21}
\end{eqnarray}
\begin{eqnarray}
\begin{array}{rcl}
r_1 &=& 2 + 3|z| + 2|z|^2 + |z|^3 ;\\
r_2 &=& 1+3|z|+3|z|^2+4|z|^3+3|z|^4+|z|^5+|z|^6; \\
r_3 &=& 5+5|z|+3|z|^2+3|z|^3.
\end{array}
\label{e21a}
\end{eqnarray}
For $|z|$ small the right-hand side is approximately $|z|/2$. If both
vectors are copied equally well, then there is a minimum value to the
copying error. The right-hand side of Eq.(\ref{e21}) takes its maximum value
approximately equal to $0.125$ when $z\simeq 0.553$.

\section{GENERAL BOUND}

Taking into account, that
\begin{eqnarray}
0\leq X_i^2\leq 1; {\mbox~~ {\rm and}~~} 0\leq |z|^2\leq 1  \label{e22a}
\end{eqnarray}
we can simplify the inequality in Eq. (\ref{e16a}), i.e.,
\begin{eqnarray}
|z|\leq |z|^2 + X_1 + X_2 + X_1 X_2.  \label{e22}
\end{eqnarray}
This allows us to go beyond specific cases and to derive a general result.

We shall adopt the quantity $X_1 + X_2$ as a measure of the total error made
in copying the two states $| s_1\rangle$ and $| s_2\rangle$. The copies are
perfect if $X_1 + X_2 =0$ and become progressively worse as its value
increase. Solving Eq.(\ref{e22}) for $X_2$ we find
\begin{eqnarray}
X_2\geq \frac{|z|(1-|z|)- X_1}{1+ X_1},  \label{e23}
\end{eqnarray}
which implies that
\begin{eqnarray}
X_1+ X_2\geq \frac{|z|(1-|z|)+ X_1^2}{1+ X_1}.	\label{e24}
\end{eqnarray}
Minimizing the right-hand side with respect to $X_1$ we find that
\begin{eqnarray}
X_1+ X_2\geq 2\left\{\left[1+|z|(1-|z|)\right]^{1/2}-1\right\}.  \label{e25}
\end{eqnarray}

A general quantum copying machine will have to copy pairs of vectors with
all values of $|z|$. In particular, it will have to copy two vectors for
which $|z|=1/2$, a value which maximizes the right-hand side of Eq.(\ref{e25}%
). For such a pair of vectors we have
\begin{eqnarray}
X_1+ X_2\geq \sqrt{5}-2.  \label{e26}
\end{eqnarray}
For this to be true, it must be the case that either $X_1 \geq (\sqrt{5}-2
)/2$ or $X_2 \geq ( \sqrt{5}-2 )/2$. This means, that for a general quantum
copying machine one has to expect that for at least one vector the size of
the copying error is $(\sqrt{5}-2 )/2\simeq 0.118$.

These considerations are closely related to recent work by Fuchs and Peres
\cite{Fuchs1}. They considered the tradeoff between disturbance and
information acquisition in quantum cryptography. Alice sends a qubit to
Bob, but in between, it is intercepted by Eve. She allows it to interact
with another qubit and sends the original on to Bob. Eve wants to
disturb the qubit she sends to Bob as little as possible yet have the
qubit she keeps contain as much information about the qubit Alice sent as
possible. Fuchs and Peres found a relation between the discrepancy rate for
Bob (disturbance) and the mutual information (Eve's information gain).
In our case we consider an interaction which produces copies. That is Eve
puts into the copy machine her qubit and Alice's qubit and what emerges
are,she hopes, two reasonably good copies of Alice's original qubit. The
assumption is then that if the copies are good the disturbance will be small
and the information gain large.

\section{MULTIPLE COPIES}

Suppose that instead of making only two copies of $|s_1\rangle$ and $%
|s_2\rangle$ we want to construct a device which will produce $(n+1)$ copies
(n actual copies plus the original). We would like to find out what the
limitations on the quality of the copies are.
Let us assume the copying transformation to be
\begin{eqnarray}
| s_j\rangle_a | Q\rangle_x \rightarrow | s_j\rangle_a | s_j\rangle_{b_1}
... | s_j\rangle_{b_n} | Q_j\rangle_x  \nonumber \\
+ | \Phi_j\rangle_{ab_1...b_n x};~~~j=1,2.  \label{e27}
\end{eqnarray}
As before we let
\begin{eqnarray}
| \Gamma_j\rangle_{ab_1...b_n x}= | s_j\rangle_a | s_j\rangle_{b_1} |
s_j\rangle_{b_n} | Q_j\rangle_x,  \label{e28}
\end{eqnarray}
and assume that $\langle \Gamma_j |\Phi _j\rangle =0$ ($j=1,2$) [in what
follows we will omit state vectors subscripts indicating the modes under
consideration, instead of $| \Gamma_j\rangle_{ab_1...b_n x} $ we will write $%
| \Gamma_j\rangle$]. What we might expect is that the more copies we make,
the poorer the quality of each copy will be. This is indeed the case.

The derivations of the inequalities are similar to those in the preceding
two sections so we shall only give the results. The inequality analogous to
that in Eq.(\ref{e16a}) is
\begin{eqnarray}
|z|\leq |z|^{n+1} (1-X_{1}^{2})^{1/2}(1-X_{2}^{2})^{1/2} +X_{1}X_{2}
\nonumber \\
+(1-|z|^{2(n+1)})^{1/2}\left[ X_{1}(1-X_{2}^{2})^{1/2}
+X_{2}(1-X_{1}^{2})^{1/2}\right].  \label{e29}
\end{eqnarray}
To analyze the multiple-copy inequalities in a transparent way, we take into
account Eq.(\ref{e22a}) and we simplify Eq.(\ref{e29}) to obtain
\begin{eqnarray}
|z|\leq |z|^{n+1}\, \left( 1-X_1^2\right)^{1/2} \left( 1-X_2^2\right)^{1/2}
\nonumber \\
+ X_1 + X_2 + X_1 X_2.	\label{e31}
\end{eqnarray}
It is useful to look at this last result in the case $X_1 = X_2 = X$. Then
one finds that
\begin{eqnarray}
X\geq \frac{\left[ 1+ \left(1-|z|^{n+1}\right)\left(|z|-|z|^{n+1}\right)
\right]^{1/2}-1 } {1-|z|^{n+1}}\equiv X_{min}.	\label{e32}
\end{eqnarray}
The right-hand side is plotted as a function of $|z|$ for several different
values of $n$ in Fig.1. One sees that $X_{min}$ is equal to zero
for $|z|=0$ and $|z|=1$ for arbitrary $n\geq 1$. This is not surprising
because we know that two mutually orthogonal states ($|z|=0$) can be copied
perfectly as many times as we wish. The case $|z|=1$ is essentially trivial,
because here the two states $|s_1\rangle$ and $| s_2\rangle$ are up to a
phase factor equal, so we are dealing with only one state. What we also see
from the figure is that for a given value of $|z|$ the bound $X_{min}$
increases as a function of $n$, that is
\begin{eqnarray}
\left.\frac{\partial X_{min}}{\partial n} \right|_{|z|=const} \geq 0.
\label{e32a}
\end{eqnarray}
This relation represents the tradeoff between the number of copies and the
noise induced by the copying procedure, i.e. the larger the number of copies
the larger the noise. Fig.~\ref{fig1} also reveals a striking asymmetry with
respect to the point $|z|=1/2$ of $X_{min}$ as a function of $|z|$. We see
that the maximum value of the function $X(|z|)$ shifts towards $|z|=1$ as $n$
increases.
\begin{minipage}{3.375in}
\begin{figure}
\begin{center}
\setlength{\unitlength}{1.cm}
\begin{picture}(6,6)
\put(-1.2,-0.8){\epsfig{file=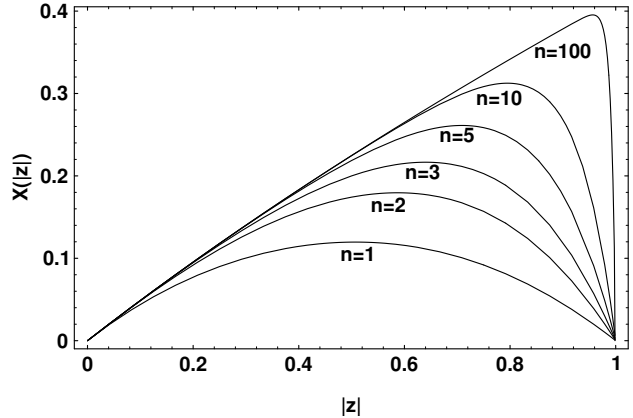,width=8.5cm}}
\end{picture}
\end{center}
\caption{ We plot the right-hand side of Eq.(32) as a function of $|z|$ for
various values of $n$ ($n=1,2,3,5,10$ and $100$). }
\end{figure}
\vskip0.3cm
\end{minipage}
 Simultaneously the maximum value increases as well and in the
limit of large $n$ is approximately equal $0.41$. It is also interesting to
note, that for $|z|$ small (when the states $|s_1\rangle$ and $| s_2\rangle$
are almost orthogonal) then
\begin{eqnarray}
X_{min}(|z|)\simeq \epsilon/2,	\label{e32b}
\end{eqnarray}
where we put $|z|=\epsilon$ ($\epsilon\ll 1$). The relation (\ref{e32b})
represents the fact that the noise induced by copying of states which are
almost orthogonal does not depend on the number of copies produced. On the
contrary, if we assume that $|z|=1-\epsilon$ (i.e. copying of states which
are almost equal), then
\begin{eqnarray}
X_{min}(|z|)\simeq n \epsilon/2,  \label{e32c}
\end{eqnarray}
which means that in the multiple-copy production of {\em almost} identical
states the error increases linearly as a function of the number of copies.

Let us briefly see what happens when $X_{1}=0$, i.\ e.\ $|s_{1}\rangle$ is
duplicated perfectly. In the limit $n\rightarrow \infty$ with $|z|<1$ we
find that $X_{2}\geq |z|$, but if $|z|=1$, then the lower bound for $X_{2}$
is zero for all $n$. For $n$ large but finite, the lower bound is
approximately equal to $|z|$  except for a region near $|z|=1$ where it
drops sharply to zero.

Finally, let us examine the $(n+1)$-copy version of Eq.(\ref{e22}). We find
\begin{eqnarray}
|z|\leq |z|^{n+1} + X_1 + X_2 + X_1 X_2,  \label{e33}
\end{eqnarray}
which implies that
\begin{eqnarray}
X_1+ X_2\geq 2\left\{\left[ 1+ |z| - |z|^{n+1}\right]^{1/2} - 1\right\}.
\label{e34}
\end{eqnarray}
The right-hand side achieves it maximum value, which is
\begin{eqnarray}
2\left\{\left[ 1+ \left(\frac{1}{n+1}\right)^{1/n} \left(\frac{n}{n+1}%
\right) \right]^{1/2} - 1\right\},  \label{e35}
\end{eqnarray}
when $|z|=(n+1)^{-1/n}$. This is an increasing function of $n$ and for large
$n$ goes to the value $2(\sqrt{2}-1)\simeq 0.83$. This implies that for a
general quantum copying machine which produces simultaneously a large number
of copies of an arbitrary input state, there must be at least one input
state for which $X_1 \geq (\sqrt{2}-1)\simeq 0.41$.

Thus we see that for a quantum copy machine which only copies two vectors or
for one which copies arbitrary input states, the lower bound for the error
in the copies increases with the number of copies made. There is clearly a
tradeoff in number of copies made versus the quality of each copy.

\section{CONCLUSION}

The unitarity of quantum mechanical transformations has allowed us to place
limits on how well quantum states can be copied. We do not know if these
limits can be realized. For example, the two quantum copy machines which
were studied in our previous paper \cite{Buzek1},
which we called the Wootters-Zurek
machine and the universal quantum copy machine, introduce more than the
minimum amount of noise into the copies they make. Finding a quantum copying
transformation which comes closest to achieving the noise limits which were
derived here is an open problem. Another problem, which we have not
addressed in the present paper is how much information is actually
transferred to the output (copy/copies and original) states.
We hope to address this problem in a future publication.

Our results can also be used to find noise limits in more general kinds of
quantum copying problems. When assessing the performance of a quantum copy
machine one needs to know not only which states are to be copied, but how
often it will be necessary to copy each one. For example, in the case where
the states $|s_{1}\rangle$ and $|s_{2}\rangle$ are to be copied, if we need
to copy $|s_{1} \rangle$ more often than $|s_{2}\rangle$, it would be better
to use a copy machine which is less noisy for $|s_{1}\rangle$ than for $%
|s_{2}\rangle$. This would result in less noise in the output, on average,
than if one were to use a copy machine which copies both states equally
well. The bounds presented in the preceding sections can be used to place
lower limits on the average amount of noise in the output for this kind of
situation.

Finally, the analysis here reveals that the feature of qubits which makes it
impossible to copy them, in general, is the fact that different qubits need
not be orthogonal. Classical information consists of bits, each of which is
in one of two completely distinguishable, and therefore orthogonal, states.
Classical information can be copied. Quantum information consists of qubits
each of which can be in any superposition of the two basis states. This
implies that two different qubits can have a nonzero inner product and are,
consequently, not completely distinguishable. It is this basic difference
between quantum and classical information which is responsible for their
different copying properties.

{\bf Acknowledgements}\newline
This work was supported by the National Science Foundation under grants INT
9221716, by the grant agency VEGA of the Slovak Academy of Sciences
(grant n. 2/1152/96), and by the United Kingdom Engineering and Physical
Sciences  Research Council.

\end{multicols}


\begin{references}
\bibitem{Ekert1}  A. Barenco and A.K. Ekert, {\em Acta Phys. Slov.} {\bf 45},
	  205 (1995).

\bibitem{Wootters1}  {\ W.K. Wootters} and {\ W.H. Zurek}, {\em Nature}
	  {\bf 299}, 802 (1982).

\bibitem{Diekes} {D. Diekes}, {\em Phys. Lett. A} {\bf 92}, 271 (1982).

\bibitem{Barnum} {H. Barnum, C.M. Caves, C.A. Fuchs, R. Josza}, and
	  {B. Schumaker}, {\em Phys. Rev. Lett.} {\bf 76}, 2818 (1996).


\bibitem{Ekert2}  A.K. Ekert, {\em Phys. Rev. Lett.} {\bf 67}, 661 (1991).

\bibitem{Bennett1}  C.H. Bennett, {\em Phys. Rev. Lett.} {\bf 68}, 3121
	  (1992).

\bibitem{Gisin} N. Gisin and B. Huttner: ``Quantum cloning, eavesdropping,
	  and Bell's inequality'', {\em Los Alamos e-print archive}
	  quant-ph/9611041 (1996).

\bibitem{DiVincenzo} {D.P. DiVincenzo}, {\em Science} {\bf 279}, 255 (1995).

\bibitem{Buzek1}  V. Bu\v{z}ek and M. Hillery, {\em Phys. Rev. A} {\bf 54},
	  1844 (1996); see also
	  V. Bu\v{z}ek, V. Vedral, M. Plenio, P.L. Knight, and M. Hillery:
	  ``Broadcasting of entanglement via local copying'',
	  {\em Los Alamos e-print archive} quant-ph/9701028 (1997),
	  to appear  in {\em Phys. Rev. A} {\bf 55} (1997).

\bibitem{Fuchs1}  C.A. Fuchs and A. Peres, {\em Phys. Rev. A} {\bf 53}, 2038
	  (1996); see also
	  C.A. Fuchs: ``Information gain vs. state disturbance in
	  quantum theory'',  {\em Los Alamos e-print archive}
	  quant-ph/9611010 (1996); and
	  C.A. Fuchs: ``Distinguishability and Accessible Information in
	  Quantum Theory'' (PhD thesis, University of New Mexico, 1995),
	  {\em Los Alamos e-print archive}
	  quant-ph/9601020 (1996).

\end{references}
\end{document}